# Towards the Realization of Higher Connectivity in MgB$_2$ Conductors: *In-situ* or Sintered *Ex-situ*?

Akiyasu Yamamoto[1,3]*, Hiroya Tanaka[1], Jun-ichi Shimoyama[1], Hiraku Ogino[1], Kohji Kishio[1] and Teruo Matsushita[2]

[1]Department of Applied Chemistry, The University of Tokyo, 7-3-1 Hongo, Bunkyo, Tokyo 113-8656, Japan

[2]Faculty of Computer Science and Systems Engineering, Kyushu Institute of Technology, 680-4 Kawazu, Iizuka, Fukuoka 820-8502, Japan

[3]Japan Science and Technology Agency, PRESTO, 4-1-8 Honcho Kawaguchi, Saitama 332-0012, Japan

*E-mail address: yamamoto@appchem.t.u-tokyo.ac.jp

**Abstract**

The two most common types of MgB$_2$ conductor fabrication technique - *in-situ* and *ex-situ* - show increasing conflicts concerning the connectivity, an effective current-carrying cross-sectional area. An *in-situ* reaction yields a strong intergrain coupling with a low packing factor, while an *ex-situ* process using pre-reacted MgB$_2$ yields tightly packed grains, however, their coupling is much weaker. We studied the normal-state resistivity and microstructure of *ex-situ* MgB$_2$ bulks synthesized with varied heating conditions under ambient pressure. The samples heated at moderately high temperatures of ~900°C for a long period showed an increased packing factor, a larger intergrain contact area and a significantly decreased resistivity, all of which indicate the solid-state self-sintering of MgB$_2$. Consequently the connectivity of the sintered *ex-situ* samples exceeded the typical connectivity range 5-15% of the *in-situ* samples. Our results show self-sintering develops the superior connectivity potential of *ex-situ* MgB$_2$, though its intergrain coupling is not yet fulfilled, to provide a strong possibility of twice or even much higher connectivity in optimally sintered *ex-situ* MgB$_2$ than in *in-situ* MgB$_2$.

# 1. Introduction

One of the distinct characteristics of $MgB_2$ among the high-temperature-superconductors (HTSs) is its conventional metallic superconductivity, *i.e.*, *s*-wave symmetry of pairing, high carrier density, long and rather isotropic coherence length, together with the high critical temperature $T_c$=40 K and high upper critical field $B_{c2}$>50 T [1]. These characteristics bring in a strongly linked current flow in randomly oriented polycrystals [2] and an easy fabrication of long length wires by the common power-in-tube (PIT) method. Additionally, being a simple intermetallic line compound from two light elements, Mg and B, and an inexpensive material costs push $MgB_2$ to a strong candidate for next-generation superconducting materials to be operated at liquid-helium-free temperatures of 15-20 K.

The reported values of critical current density $J_c$ at 20 K for $MgB_2$ bulks, wires and tapes $10^5$–$10^6$ $Acm^{-2}$ [3-5] turned out to be apparently lower than the depairing current density, $J_d$(20 K)~$B_c/\mu_0\lambda$~$10^8$ $Acm^{-2}$, where $B_c$ is the thermodynamic critical field, $\mu_0$ is the permeability of vacuum and $\lambda$ is the penetration depth. Indeed very high $J_c$ values reaching $10^7$ $Acm^{-2}$ at 20 K have been reported for epitaxial thin films [6,7]. Grain boundaries work as predominant flux pinning centers in $MgB_2$ and the doping of carbon-based compounds, such as graphite [8,9], $B_4C$ [10], SiC [11], and organic compounds [12-14] and low-temperature synthesis [15] are reported to be effective in increasing $J_c$. The degradation of crystallinity, *i.e.*, the distortion of honeycomb boron lattice, is believed to be the origin of the enhancement of flux pinning for both cases and particularly contributes to the improvement of $J_c$ under high magnetic fields [16,17].

A reduction in the effective current-carrying cross-sectional area of the sample was suggested by Rowell to explain the large gap of $J_c$ between films and wires [18]. The Josephson junction model of the grain boundaries [18], the two-band model [19,20], the anisotropy model [21], and the oxide barrier model [22] were considered to affect the limited transport properties of $MgB_2$. In our previous study, we applied a mean-field theory to the three-dimensional percolation problem to understand the anomaly suppressed connectivity in rather weak-link-free $MgB_2$ polycrystals [23,24]. The mean-field theory quantitatively showed that the packing factor (*P*) of polycrystals, impurity layers at grain boundaries, and anisotropy are the limiting factors of the connectivity.

*In-situ* and *ex-situ* methods have been developed to manufacture $MgB_2$ bulks, wires, and tapes. Perhaps the most commonly studied method is the *in-situ* method, that is the formation of $MgB_2$ simply from mixed powders of Mg+2B, since a relatively high $J_c$ value is easily attained owing to its reasonably strong intergrain coupling. In an *in-situ* reaction process, Mg grains melt and diffuse into B grains, and transform into voids resulting in a low bulk density ($P\sim50\%$) and a low connectivity. On the other hand, the *ex-situ* method using prereacted $MgB_2$ powder is favorable in terms of bulk density. A packing factor close to ~75% (the close packing of spheres) can be expected. Even unsintered, as-pressed *ex-situ* $MgB_2$ tapes show a relatively high transport $J_c$ value of ~$10^4$ $Acm^{-2}$ at 20 K [25]. Heat treatment after cold working is effective in improving $J_c$ through the strengthening of intergrain coupling [3,26-28]. The $J_c$ of heat treated *ex-situ* $MgB_2$ is, however, generally lower than that of *in-situ* $MgB_2$, likely due to the fact that intergrain coupling is insufficient compared with the strong coupling in the *in-situ* $MgB_2$. The connectivity of reasonably high $J_c$ *ex-situ* $MgB_2$ tapes is reported to be less than 10% [29,30], which is obviously lower than the typical connectivity of *in-situ* $MgB_2$, 5-15% [23,31]. Since the packing factor of *ex-situ* $MgB_2$ is higher than that of *in-situ* $MgB_2$, a better connectivity, even higher than that of *in-situ*, is naturally expected *if a strong intergrain coupling is achieved*.

In this paper we carefully investigated the microstructure, normal-state resistivity and electrical connectivity of *ex-situ* $MgB_2$ polycrystalline bulks prepared using systematically varied heating conditions under ambient pressure. In particular we employed long heat treatments at high temperatures of ~900°C to promote the self-sintering of $MgB_2$ grains. In order to prevent the decomposition of $MgB_2$ by the vaporization of Mg at high temperatures, prereacted $MgB_2$ powders were sealed and heated in a metal sheath using our powder-in-closed-tube (PICT) technique [32]. We observed evidence for the solid-state self-sintering of $MgB_2$ and its strong effect on the enhancement of connectivity. On the basis of the results, we will compare the intergrain coupling nature of *in-situ* and *ex-situ* $MgB_2$, and discuss the prospects for further improvement of connectivity in $MgB_2$ conductors.

**2. Experimental Procedure**

*Ex-situ* $MgB_2$ polycrystalline bulk samples were fabricated by the PICT method. The detailed fabrication methods for the bulk samples can be found elsewhere [32,33]. Laboratory-made $MgB_2$ powder or commercially available $MgB_2$ powder (99% purity, several tens of microns in size, Alfa Aesar) was used as a starting material. The $MgB_2$ powder was filled into a stainless-steel (SUS316) tube, then the tube was uniaxially pressed under 500 MPa with both ends closed by mechanical pressing. Finally each tube was heated at 750-950°C for 3-96 h in an evacuated quartz ampoule. Laboratory-made $MgB_2$ powder was prepared by grinding the *in-situ*-processed bulk (heating condition: 900°C for 2 h) synthesized from mixed powders of Mg (99.5% purity) and B (99% purity) with the molar ratio of 1:2. *In-situ-* and *diffusion*-processed bulks were prepared from Mg and B powders for comparison.

Constituent phases of the samples were analyzed by the powder x-ray diffraction (XRD) method using Cu $K\alpha$ radiation. Microstructural observation was performed using a scanning electron microscope (SEM; JEOL JSM-7000F). The packing factor ($P$) of the samples was measured with a micrometer caliper and a weighing balance. Resistivity measurements were performed by the conventional four-point probe method with ac current of 15 Hz using a physical property measurement system (PPMS; Quantum Design PPMS Model 6000).

## 3. Results and Discussion
### 3. 1. Resistivity and connectivity

Figure 1(a) shows the electrical resistivity $\rho$ as a function of temperature for the *ex-situ* bulks from laboratory-made $MgB_2$ powder with a systematically varied degree of sintering, by heating at different temperatures from 750 to 900°C. The as-pressed $MgB_2$ bulk before heat treatment, *in-situ* bulk, and *diffusion* bulk are also shown for comparison. The as-pressed bulk has a large resistivity of $\sim 1 \times 10^4$ $\mu\Omega$cm at room temperature and shows an unusual temperature dependence with a very slight upturn below $\sim 100$ K, resulting in low $RRR = \rho(300\text{ K})/\rho(40\text{ K}) = 1.1$. The resistivity of the as-pressed bulk does start dropping at $\sim 39$ K and reaches zero resistance; however, its superconducting transition is broad with $\Delta T_c > 10$ K [Fig. 1(b)]. The high $\rho$, small $RRR$, and large $\Delta T_c$ suggest that intergrain coupling is weak. On the other hand, the *ex-situ* bulks heat-treated at above 850°C show a successively lower resistivity as the heat treatment temperature increases, indicative of evolution of intergrain coupling by sintering. Indeed we observed an increase in packing factor for the heat-treated bulks compared to the as-pressed bulk. The bulk heated at 900°C for 48 h, with the highest degree of sintering, shows resistivities of 50 $\mu\Omega$cm at 300 K and 15 $\mu\Omega$cm at 40 K, which are 2 or 3 orders of magnitude lower than that of the as-pressed bulk and even lower than that of the typical *in-situ* bulk. Consequently the resistive transition becomes sharper with the progression of sintering, and the bulks sintered above 850°C show small $\Delta T_c < 1$ K which is comparable to that of the *in-situ* bulks.

The evolution of transport current connectivity by sintering is summarized in Fig. 2. The zero resistance temperature $T_{R0}$ is defined by $\rho < 10^{-1}$ $\mu\Omega$cm. The electrical connectivity $K = \Delta\rho_g/\Delta\rho$, where $\Delta\rho_g = \rho_g(300\text{ K}) - \rho_g(40\text{ K}) \equiv 6.32$ $\mu\Omega$cm [23] and $\Delta\rho = \rho(300\text{ K}) - \rho(40\text{ K})$ are the difference in resistivity of the ideal $MgB_2$ grains and that of a sample, respectively, is plotted as a function of sintering temperature. Both $T_{R0}$ and *K* show a rapid increase above 850°C, and the maximum connectivity is obtained for the bulk sintered at 900°C. Sintering above 950°C reduces connectivity, probably due to the decomposition of $MgB_2$ as evidenced by B-rich impurity phases observed by compositional analyses (not shown here). The prolonged heat treatment further promoted sintering and improvement in connectivity. The bulk sintered at 900°C for 48 h shows *K*~18% which is among the highest values for $MgB_2$ polycrystals except samples synthesized by *diffusion* process [23,34] or under high pressure [35]. Here the connectivity of the sintered *ex-situ* bulks exceeds the typical range of *in-situ* processed bulks and wires which is 5-15%.

### 3. 2. Microstructure

Figure 3 summarizes typical microstructural features of $MgB_2$ polycrystalline bulks prepared by *in-situ* [Fig. 3(a)] and *ex-situ* [Fig. 3(b)] methods. Here, gray, black, and white contrasts in the secondary electron images correspond to $MgB_2$ grains, pores, and impurity phases, such as MgO, respectively. The *in-situ*-processed sample shows a porous microstructure with large voids typically 10-50 μm in size. Spaces filled with Mg powders before the heat treatment transform into voids through the reaction with B. On the other hand, a characteristic microstructure different from that of the *in-situ* bulk can be seen in the *ex-situ* bulk sintered at 900°C for 24 h. In Fig. 3(b) islands of $MgB_2$ grains and particles with a size of ~10 μm are dispersed and the voids occupy the gap between the islands. For the *ex-situ* bulk, the shape of the voids is apparently different from that of the *in-situ* bulk, and their size is much smaller (typically less than 10 μm). One can see that the intergrain coupling between $MgB_2$ grains/particles is poor in contrast to that in the *in-situ* sample where a strongly linked network of $MgB_2$ grains is observed. The weak intergrain coupling is considered to be the reason for the rather restricted *K* observed in the *ex-situ* bulk [Fig. 1(a)] though the packing factor of the *ex-situ* bulk (64%) is much higher than that of the *in-situ* bulk (48%).

Higher magnification images of the polished cross-sectional surface of $MgB_2$ bulks are shown in Fig. 4 to manifest intergrain coupling between $MgB_2$ grains. The as-pressed $MgB_2$ bulk before sintering shows fine $MgB_2$ grains/particles are tightly packed and neither of intergrain or grain-particle coupling can be seen [Fig. 4(b)]. After heating at 900°C, the surface area of $MgB_2$ grains decreased and the size of voids increased compared with those observed in the as-pressed and coupling between $MgB_2$ grains/particles were also clearly observed [Fig. 4(c)], all of these suggest that *solid-state* self-sintering occurred during the heat treatment. On this magnification scale, we did not observe impurity phases or cracks at grain boundaries of the sintered *ex-situ* $MgB_2$ bulk. Though the area of coupling between grains in the sintered *ex-situ* bulk is smaller than that in the *in-situ* bulk [Fig. 4(a)] such coupling is believed to contribute significantly as an effective path for the transport current in *both* normal and superconducting states.

## 4. Discussion

It is well known that the $J_c$ of the *ex-situ* $MgB_2$ can be largely enhanced by heat treatment. The improvement of intergrain coupling by sintering or the removal of volatile impurities from grain boundaries can be considered as the reason. However, there are few reports on the self-sintering of $MgB_2$. Dancer *et al.* studied the effects of a range of heat treatment (widely varied from 200 to 1100°C for 1 h) on the packing factor and the amount of the MgO impurity phase for *ex-situ* $MgB_2$ bulks [36]. After heat treatments they observed little sign of sintering even at 1100°C and a small change (<3%) in packing factor. Our bulk samples showed a partially sintered microstructure which is similar to that of spark-plasma-sintered (SPS) [37] or high-pressure-processed [35,38] bulks together with an approximately 10% increase in packing factor. The lowered resistivity by the orders of magnitude observed in the *ex-situ* bulks suggests that solid-state self-sintering occurs with a long-period heat treatment at high temperatures of ~900°C.

Thermodynamically decomposition of $MgB_2$ takes place under a low Mg partial pressure [39] as

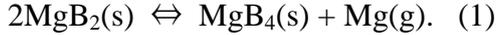
$$2MgB_2(s) \Leftrightarrow MgB_4(s) + Mg(g). \quad (1)$$

Such a decomposition of $MgB_2$ was experimentally observed by the loss of gaseous Mg at temperatures as low as ~610°C [40], and the formation of $MgB_4$ obviously causes the degradation of superconducting properties [41]. In our case, we heated the samples in a closed system, *i.e.*, a sealed stainless-steel tube, using the PICT method and precisely controlled the amount of Mg by preventing vaporization. A significant reaction between vaporized Mg and quartz ampoule occurred when the stainless-steel tube was not sealed. By using the PICT technique, we did not observe a trace of B-rich phases in the sintered bulks except samples heated for a very long period where the reaction between the stainless-steel sheath and Mg occurred. Therefore the sintering of $MgB_2$ should be performed under precisely controlled conditions of highly reactive, volatile Mg.

We show in Fig. 5 the relationship between packing factor $P$ and the connectivity $K$ for *ex-situ* bulks synthesized from both commercial and laboratory-made powders and heated at 900°C for different periods and *in-situ* bulks [23]. A higher connectivity was observed with an increase in $P$ through sintering, and a trend between $P$ and $K$ was observed for the *ex-situ* bulks. Note that the data for the *ex-situ* bulks shown in Fig. 5 are some of the well-connected samples which exclude those of samples showing indications of B-rich phase formation. $K$ for the samples with insufficient sintering or with impurity phases scatters below such a trend. For *in-situ* $MgB_2$, the relationship between $P$ and $K$ can be well explained by the mean-field theory for a three-dimensional site percolation system according to the equation [23]

$$K = \frac{(aP)^2 - P_c^2}{1 - P_c^2}, \quad (2)$$

where $a$ is the fraction of effective $MgB_2$ grains that can carry current and $P_c$ is the critical packing factor and is 0.3117 for the three-dimensional cubic site system [42]. Interestingly, the observed *P-K* trend for the *ex-situ* bulks shifts to higher $P$ values than that for the *in-situ* bulks, suggesting that the limiting mechanisms of connectivity for the bulks from two processes are different. Suppose the three-dimensional percolation model works for the *ex-situ* bulks, the result indicates that either critical packing factor $P_c$ is higher and/or intergrain coupling between $MgB_2$ grains/particles (which corresponds to $a$) is weaker in the sintered *ex-situ* bulks. Considering that the contacted area between $MgB_2$ grains in the *ex-situ* bulks is limited by the porosity gaps [Figs. 3(b), 4(c)], it is considered that intergrain coupling is still insufficient compared with that in the *in-situ* bulks.

Thus far the connectivity of *ex-situ* $MgB_2$ is a trade-off balance between the higher packing factor and the weaker intergrain coupling. Equation (2) predicts a high connectivity of 30-40% for moderately sintered *ex-situ* $MgB_2$ with $P$~75% owing to its large $P$. Just a ~25% increase in $P$ compared with that in *in-situ* $MgB_2$ results in twice or three times higher connectivity in *ex-situ* $MgB_2$, if a sufficient arrangement of surface contact between grains is achieved. Our results suggest that under controlled atmosphere of Mg pressure, the solid-state self-sintering of $MgB_2$ occurs and significantly improves the intergrain coupling by heat treatment under an ambient pressure. Given that homogeneous, single starting powder is favorable for the fabrication of wires by the PIT method, sintered *ex-situ* $MgB_2$ has advantages in both connectivity and long-length wire fabrication. We believe the issues on the microstructure of sintered *ex-situ* $MgB_2$ bulks, such as large agglomerates and gaps between grains, can reasonably be solved by the optimization of powder preparation and heat treatment conditions in near future.

Finally we briefly mention the $J_c$ of sintered *ex-situ* $MgB_2$. What surprised us is that the long heat treatment did not yield a significant increase in grain size. Indeed the sintering promoted agglomerate formation; however, it just enhanced surface contact and did not promote grain growth [Fig. 4(c)]. This is in strong contrast to that observed in *in-situ* MgB$_2$ bulks heated at high temperatures for a long period where grain growth occurred and a marked deterioration in $J_c$ was observed. We observed a higher $J_c$ value in the sintered *ex-situ* bulks than in the optimized *in-situ* bulks. Such critical current properties in the relationship between the connectivity and microstructure of the sintered *ex-situ* bulks will be reported in detail in a subsequent paper [33].

## 5. Conclusions

We studied the normal-state resistivity and microstructure of *ex-situ* MgB$_2$ bulks synthesized with varied heating conditions under ambient pressure. The samples heated at moderately high temperatures of ~900°C for a long period showed an increased packing factor, a larger intergrain contact area, and a significantly decreased resistivity, all of which indicate the solid-state self-sintering of MgB$_2$. Consequently the connectivity of the sintered *ex-situ* samples exceeded the typical connectivity range 5-15% of the *in-situ* samples. Our results show self-sintering can develop the superior connectivity potential of *ex-situ* MgB$_2$, though its intergrain coupling is not yet fulfilled, to provide a strong possibility of realizing twice or even much higher connectivity in optimally sintered *ex-situ* MgB$_2$ than in *in-situ* MgB$_2$.

**Acknowledgements**

This work was partially supported by Grants-in-Aid for Scientific Research from the Japan Society for the Promotion of Science Nos. 23246110 and 22860019.

**Figure captions**

**Fig. 1.** (a) Temperature dependence of resistivity for the *ex-situ* MgB$_2$ bulks sintered at different temperatures. *Ex-situ* bulk without heat treatment (as-pressed); *in-situ-* and *diffusion*-processed bulks [23] are also shown for comparison. Figure 1(b) manifests resistive transitions near $T_c$.

**Fig. 2.** Zero resistance temperature $T_{R0}$ and connectivity $K$ as a function of sintering temperature for the *ex-situ* MgB$_2$ bulks. All the samples were sintered for 24 h except the samples without heat treatment and sintered for 48 h.

**Fig. 3.** Secondary electron images for the polished surface of MgB$_2$ polycrystalline bulks. (a) *In-situ*-processed MgB$_2$ bulk from Mg and B, and (b) *ex-situ*-processed MgB$_2$ bulk (sintered at 900°C for 24 h) from laboratory-made MgB$_2$ powder.

**Fig. 4.** High-magnification secondary electron images for the polished surface of MgB$_2$ polycrystal bulks. (a) Dense regions of the *in-situ* bulk, (b) *ex-situ* bulk without heat treatment (as-pressed), and (c) *ex-situ* bulk sintered at 900°C for 24 h.

**Fig. 5.** Relationship between packing factor $P$ and connectivity $K$ for the *in-situ* (including *diffusion*-processed) [23] and *ex-situ* MgB$_2$ bulks.

**Figure 1**

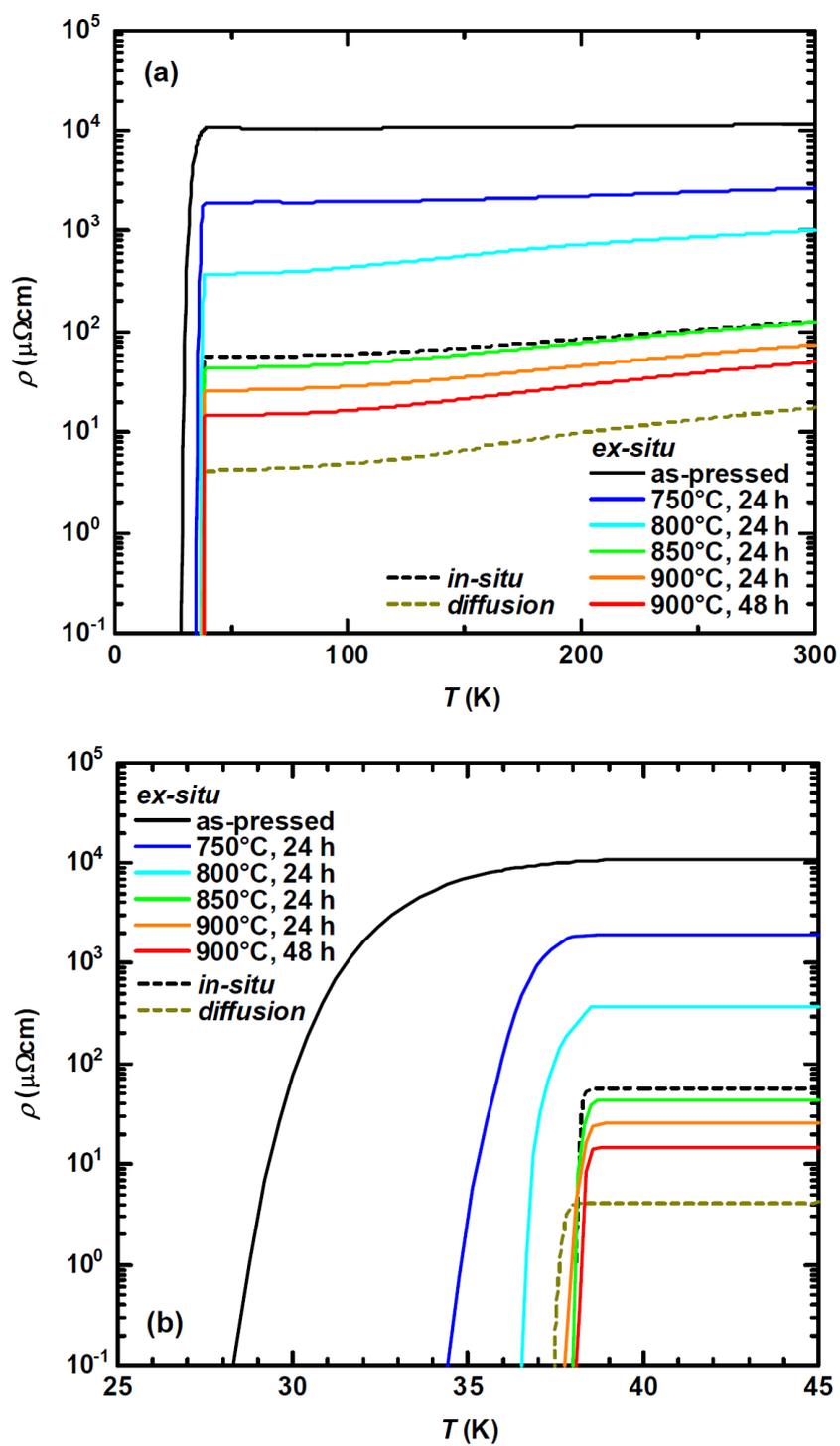

Yamamoto et al.

**Figure 2**

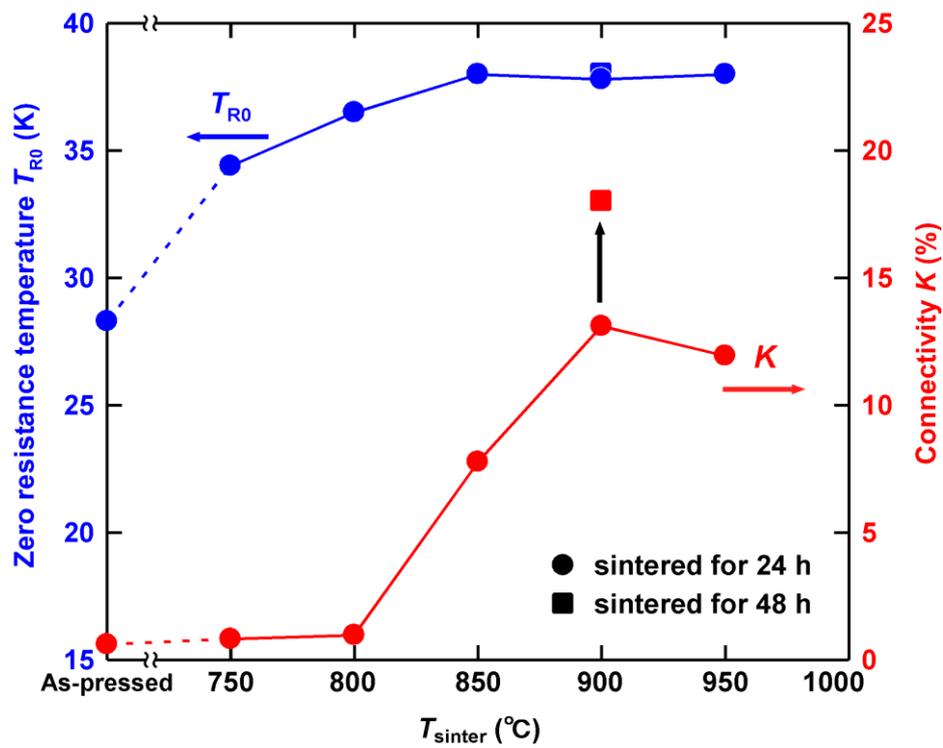

Yamamoto et al.

**Figure 3**

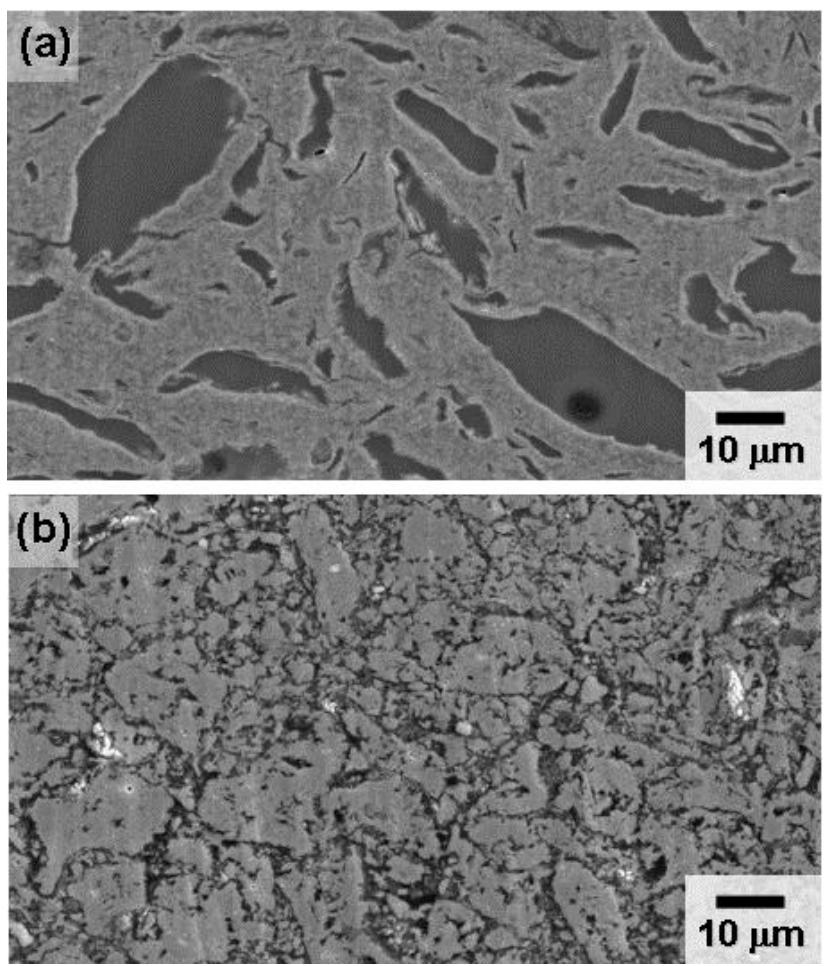

Yamamoto et al.

**Figure 4**

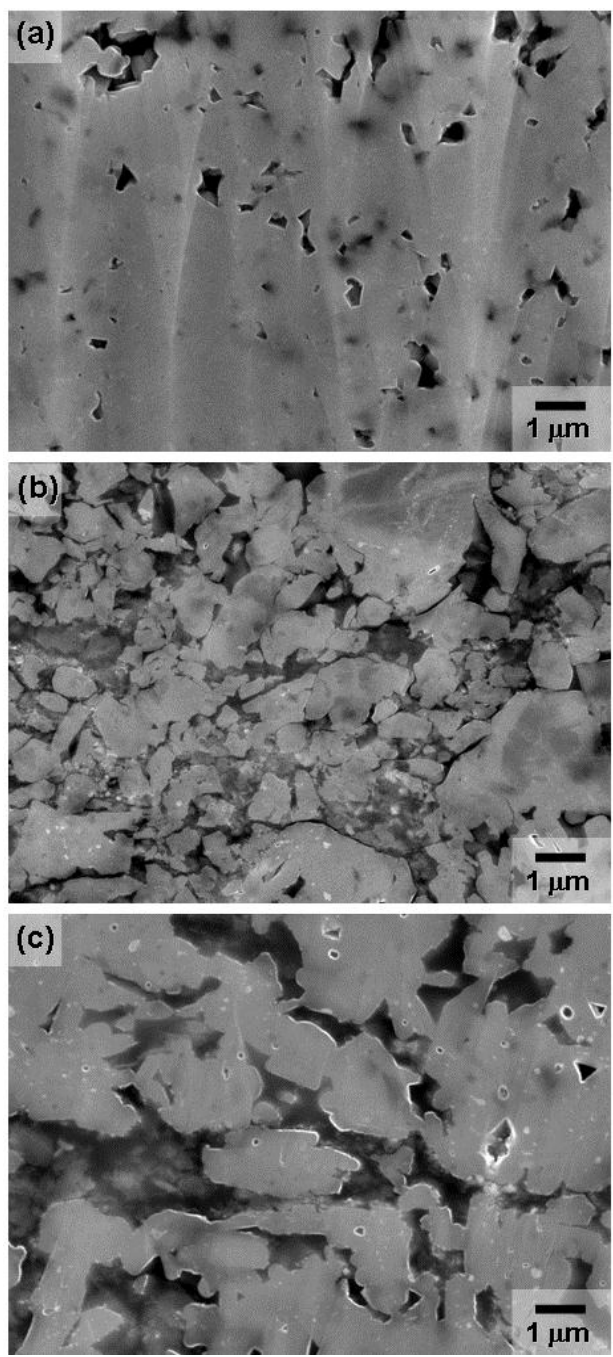

Yamamoto et al.

**Figure 5**

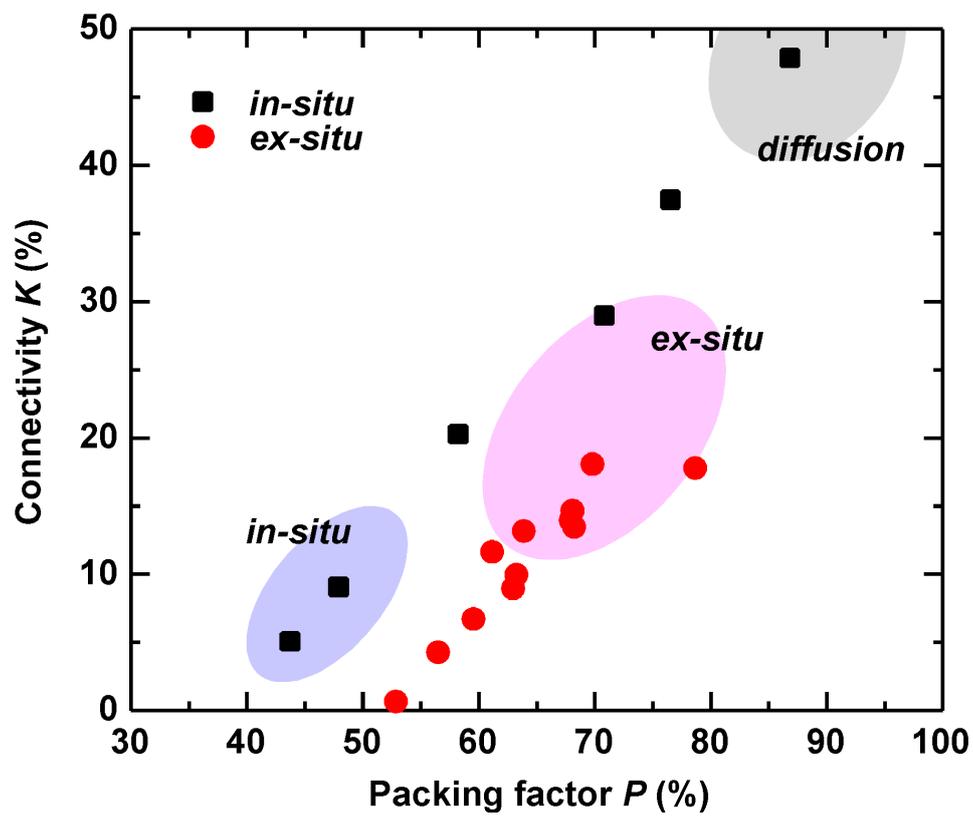

Yamamoto et al.